\documentstyle[prl,aps,psfig]{revtex}

\begin{document}
\title{Quantum vacuum fluctuations}
\author{%
Serge Reynaud~$^{\text{a}}$,\ \
Astrid Lambrecht~$^{\text{a}}$,\ 
Cyriaque Genet~$^{\text{a}}$,\ 
Marc-Thierry Jaekel~$^{\text{b}}$
}
\address{%
\begin{itemize}\labelsep=2mm\leftskip=-5mm
\item[$^{\text{a}}$]
Laboratoire Kastler Brossel \thanks{Laboratoire de l'Ecole 
Normale Su\-p\'e\-rieu\-re, de l'Uni\-ver\-si\-t\'e 
Pierre et Marie Curie et du CNRS}
\thanks{Website~: www.spectro.jussieu.fr/Vacuum}, 
UPMC case 74, Jussieu, F75252 Paris Cedex 05 \\
\item[$^{\text{b}}$]
Laboratoire de Physique Th\'eorique de l'ENS 
\thanks{Laboratoire du CNRS, de l'Ecole Normale 
Su\-p\'e\-rieu\-re et de l'Uni\-ver\-si\-t\'e Paris-Sud}, 
24 rue Lhomond, F75231 Paris Cedex 05 
\end{itemize}
}
\maketitle

\begin{abstract}{%
The existence of irreducible field fluctuations in vacuum is an important prediction 
of quantum theory. These fluctuations have many observable consequences, like
the Casimir effect which is now measured with good accuracy and agreement 
with theory, provided that the latter accounts for differences between real 
experiments and the ideal situation considered by Casimir. But the vacuum 
energy density calculated by adding field mode energies is much larger 
than the density observed around us through gravitational phenomena.
This ``vacuum catastrophe'' is one of the unsolved problems at the interface 
between quantum theory on one hand, inertial and gravitational phenomena on the 
other hand. It is however possible to put properly formulated questions in the
vicinity of this paradox. These questions are directly connected to observable 
effects bearing upon the principle of relativity of motion in quantum vacuum.
}\end{abstract}

Observation of movement forces us to conceive the notion of a space 
in which movement takes place. The absence of resistance opposed to 
movement by space then leads to the concept of empty space, or vacuum. 
These logical statements were formulated by Leucippus and Democritus more
than 2000 years ago \cite{Russell}. They raise paradoxes which have shown 
their pertinence from that time until today. 

The question of relativity of motion played an important role in the birth
of modern physics. Galileo emphasized that motion with uniform velocity 
in empty space is indistinguishable from rest but, also, that this property 
is only true when resistance of air to motion can be ignored \cite{Galileo}. 
Newton stressed this fact \cite{Newton} by identifying empty space in 
which motion takes place with the experimental vacuum which was studied 
by Pascal, von Guericke and Boyle after the pioneers Galileo and Torricelli. 
The same question played anew a key role in Einsteinian theory of relativity
which freed space from the absolute character it had in classical Newtonian 
theory \cite{EinsteinR}, in contradiction with the Galilean principle of 
relativity (named in this manner by Einstein). 

The emergence of quantum theory has profoundly altered our conception of 
empty space by forcing us to consider vacuum as the realm of quantum field 
fluctuations. This solves old paradoxical questions about the very nature
of vacuum. In quantum theory, vacuum becomes a well-defined notion and 
the question of relativity of motion is given a satisfactory understanding.
However, a new difficulty arises which involves the energy density of
vacuum fluctuations and, precisely, the contradiction between the large values 
predicted by quantum theory and the evidence of its weak, if not null, value
in the world around us. These questions are discussed in more detail below 
after a few extra historical remarks devoted to the birth of quantum theory.

The classical idealization of space as being absolutely empty was already affected
by the advent of statistical mechanics, when it was realized that space is in fact 
filled with black body radiation which exerts a pressure onto the boundaries of 
any cavity. It is precisely for explaining the properties of black body radiation 
\cite{Darrigol00} 
that Planck introduced his first quantum law in 1900 \cite{Planck00}. In modern terms, 
this law gives the energy per electromagnetic mode as the product $E$ of 
the energy of a photon $\hbar \omega \equiv h \nu $ by a number of photons $n$ per mode 
\begin{equation}
E = n \hbar \omega  \qquad \qquad 
n = \frac {1}{\exp \left( \frac{\hbar \omega}{k_{\rm B}  T} \right) - 1}
\end{equation}
Unsatisfied with his first derivation, Planck resumed his work in 1912 and derived 
a different result where the energy contained an extra term \cite{Planck12}
\begin{equation}
E = \left( \frac 12 + n \right) \hbar \omega  
\end{equation}
The difference between the two Planck laws just corresponds to what we now call 
``vacuum fluctuations''. Whereas the first law describes a cavity entirely
emptied out of radiation at the limit of zero temperature, the second law tells
us that there remain field fluctuations corresponding to half the energy of a
photon per mode at this limit. 

The story of the two Planck laws and of the discussions they raised is related
in a number of papers, for example \cite{Milonni91,Sciama91}. 
Let us recall here a few facts~: Einstein and Stern noticed in 1913 that the second law, 
in contrast with the first one, has the correct classical limit at high temperature
\cite{Einstein13}
\begin{equation}
\left( \frac 1 2 + n \right) \hbar \omega = k_{\rm B}  T + O \left( \frac 1T \right) 
\qquad \qquad T \rightarrow \infty 
\end{equation}
Debye was the first to insist on observable consequences of zero-point 
fluctuations in atomic motion, by discussing their effect on the intensities 
of diffraction peaks \cite{Debye14} whereas Mulliken gave the first experimental
proof of these consequences by studying vibrational spectra of molecules 
\cite{Mulliken24}. Nernst is credited for having first noticed that
zero-point fluctuations also exist for field modes \cite{Nernst16} which
dismisses the classical idea that absolutely empty space exists and may be 
attained by removing all matter from an enclosure and lowering the temperature 
down to zero. At this point, we may emphasize that these discussions took place
before the existence of these fluctuations was confirmed
by fully consistent quantum theoretical calculations.

We now come to a serious difficulty of quantum theory which was noticed 
by Nernst in his 1916 paper. Vacuum is permanently filled with electromagnetic 
field fluctuations propagating with the speed of light, as any free field, and
corresponding to an energy of half a photon per mode. In modern words, quantum 
vacuum is the field state where the energy of field fluctuations is minimal. 
This prevents us from using this energy to build up perpetual motions 
violating the laws of thermodynamics. However, this also leads to a serious problem
which can be named the ``vacuum catastrophe'' \cite{Adler95}, in analogy 
with the ``ultraviolet catastrophe'' which was solved by Planck in 1900
for the black body problem. When the total energy is calculated by 
adding the energies of all field modes in the vacuum state, an infinite 
value is obtained. When a high frequency cutoff $\omega_{\rm max}$ is introduced, the 
energy density is read
\begin{equation}
e = \frac {\hbar} {160 \pi^2 c^3} \left( 20 \omega_{\rm max}^4 + \theta^4 \right) 
\qquad \qquad \theta = \frac {2 \pi k_{\rm B}  T} {\hbar}
\label{eqVacuumEnergyDensity}
\end{equation}
The first term, proportional to $\omega_{\rm max}^4$, is the density of vacuum energy 
per unit volume and it diverges when $\omega_{\rm max}\rightarrow\infty$. This has to
be contrasted with the second term, the Stefan-Boltzmann energy density of 
black body fluctuations, which is finite and proportional to 
$\theta^4$, where $\theta$ is the temperature measured as a frequency. 
The problem is not a merely formal difficulty, since the calculated value is
enormously larger than the mean energy of vacuum observed in the world around 
us through gravitational phenomena. And this is true not only when $\omega_{\rm max}$ 
is chosen as the Planck frequency. In fact the problem persists for any value
of the cutoff which preserves the laws of quantum theory at the energies where
they are well tested. 

In other words, the vacuum catastrophe has constituted from 1916 until today 
a major discrepancy between the classical theory of general relativity on one
side and the quantum theory on the other side. This has crudely been stated 
by Pauli \cite{Pauli33} ``At this point it 
should be noted that it is more consistent here, in contrast to the material 
oscillator, not to introduce a zero-point energy of $\frac 12 h \nu$ per 
degree of freedom. For, on the one hand, the latter would give rise to an
infinitely large energy per unit volume due to the infinite number of degrees
of freedom, on the other hand, it would be principally unobservable since
nor can it be emitted, absorbed or scattered and hence, cannot be contained
within walls and, as is evident from experience, neither does it produce
any gravitational field'' (translation reproduced from \cite{Enz74}).

This problem, also known as the ``cosmological constant problem''
because of its obvious connection with the introduction of a cosmological
constant in Einstein gravitation equations \cite{Abbott88,Weinberg89}, 
has remained unsolved during the twentieth century, despite considerable 
efforts for proposing solutions \cite{Demianski00,Weinberg00,Witten00}.
It has the status of a paradox, lying just at the crucial interface
between quantum theory and gravity, and pointing at the necessity
of substantial reformulations in the present theoretical formalism.
We want however to emphasize in the present paper that a number of hints
are already available which correspond to well-defined questions 
with proper answers in the present formalism and, simultaneously,
observable consequences.

Certainly, we have to acknowledge with Pauli that the mean value of vacuum 
energy does not contribute to gravitation as an ordinary energy. This is
just a matter of evidence since the universe would look very differently 
otherwise. In other words, the reference level setting the zero 
of energy for gravitation theory appears to be finely tuned to fit 
the mean value of vacuum energy. However, this cannot lead to dismiss the 
effects of vacuum energy. 
As a matter of fact, even for an exact cancellation of the contribution
of mean vacuum energy, energy differences and energy fluctuations still have to 
contribute to gravitation. This point will be discussed in more detail below.

It is no more possible to uphold, as Pauli did, that vacuum fluctuations 
cannot ``be emitted, absorbed, scattered ... or contained within walls''.
Vacuum fluctuations of the electromagnetic field have well known observable 
consequences on atoms \cite{Cohen88} and, more generally, microscopic scatterers
\cite{Itzykson85}. An atom interacting only with vacuum fields suffers 
spontaneous emission processes induced by these fields. When fallen in its 
ground state, the atom can no longer emit photons but its coupling to vacuum
still results in measurable effects like the Lamb shift of absorption frequencies. 
Two atoms located at different locations in vacuum experience an attractive 
Van der Waals force. While studying this effect which plays an important role 
in physico-chemical processes, Casimir discovered in 1948 that a force, now named 
after him, arises between two mirrors placed in vacuum and is then a macroscopic 
analog of the microscopic Van der Waals forces \cite{Sparnaay89}. 
This effect is discussed below. 

We also want to recall that vacuum fluctuations of electromagnetic fields
are, in a sense, directly detected through the study of photon noise
in quantum optics. In simple words, photon noise reflects field fluctuations.
This idea was first formulated by Einstein during his early efforts to build
up a consistent theory of light \cite{Einstein05Q,Einstein09,Wolf79}.
Here we will illustrate it in a simple manner by discussing in the words of 
modern quantum optics a gedanken experiment where an electromagnetic field
impinges a beam splitter \cite{Reynaud90}. 
\begin{figure}[h]
\centerline{\psfig{figure=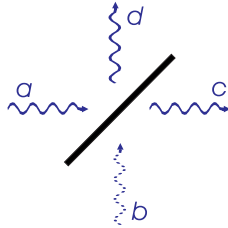,height=3cm}}
\caption{Schematic description of a simple quantum optical experiment~: a beam of
photons enters the input port `a' of a beam splitter; photon noise is analyzed 
through intensity fluctuations of the beams in the output ports `c' and `d';
its statistics can be understood as determined by the field fluctuations 
entering the usually disregarded input port `b'.}
\label{figSplitter}
\end{figure}
This experiment can be analyzed in terms of photons, each one having
probabilities of 50 \% for being either transmitted or reflected.
The random character of this process results in the photon noise
observed behind the beam splitter as fluctuations of the number of 
photons in the output ports. 
If we now think in terms of electromagnetic fields, we first represent
the field ${\cal E}$ in any of the involved modes, with a frequency $\omega$, as a 
sum over two quadrature components ${\cal E}_1$ and ${\cal E}_2$ obeying Heisenberg inequality 
\begin{equation}
{\cal E} = {\cal E}_1 \cos \omega t +{\cal E}_2 \sin \omega t \qquad \qquad \Delta {\cal E}_1 
\Delta {\cal E}_2 \geq {\cal E}_0^2 
\end{equation}
This is basically the reason for the necessity of quantum fluctuations of 
electromagnetic fields, ${\cal E}_0$ being a constant characterizing the level
of vacuum fluctuations. The vacuum state corresponds to the case where 
fields have a null mean value and a minimum energy, which leads
to $\Delta {\cal E}_1^2 = \Delta {\cal E}_2^2 = {\cal E}_0^2 $. But Heisenberg 
inequality allows one to obtain ``squeezed'' states with a noise 
on a given quadrature component smaller than in vacuum \cite{Reynaud92}. 

Simple calculations then lead to the result that number fluctuations in one of
the output ports are determined by the field fluctuations corresponding to one 
of the quadrature components, say ${\cal B}_1 $, in the usually disregarded input 
port `b' (see Figure \ref{figSplitter}) 
\begin{equation}
\delta n = \delta n_c - \delta n_d \sim \left| {\cal A} \right| \delta {\cal B}_1 
\qquad \qquad \Delta n^2 \sim \left| {\cal A} \right|^2  \Delta {\cal B}_1^2 
\end{equation}
Usually the field fluctuations entering the port `b' are in the vacuum state, 
so that the variance $\Delta {\cal B}_1^2$ is just the constant ${\cal E}_0^2 $ which 
leads to the ordinary Poissonian statistics of photon noise 
$\Delta n^2 = \left\langle n_a \right\rangle$. 
But this statistics may be modified by manipulating the fluctuations of ${\cal B}_1$. 
In particular, letting a squeezed field with $\Delta {\cal B}_1^2 < {\cal E}_0^2 $ 
enter the input port `b' allows one to obtain a sub-Poissonian statistics on
the variable $n$. 

This means that photon noise merely reflects quantum field fluctuations 
entering the optical system and can therefore be controlled by manipulating
the components responsible for fluctuations of the signal of interest.
For the experiment sketched on Figure \ref{figSplitter}, the random scattering 
of photons from the input beam `a' into the two output beams `c' and `d' 
may be made more regular by squeezing the fluctuations of ${\cal B}_1$.
More generally, this idea may be used for improving the sensitivity of 
ultra-high performance optical sensors such as the interferometers designed 
for gravitational wave detection \cite{Jaekel90}. It may be applied to a
measurement apparatus as sophisticated as the cold damping accelerometer 
developed at ONERA for testing the equivalence principle in space experiments. 
In this case, fluctuations entering all electrical or mechanical ports have 
to be taken into account, including ports involved in active elements 
\cite{Grassia00}. The analysis can be pushed further, up to the problem of 
a satellite actively controlled to follow a dragfree geodesic trajectory. 
In this situation of interest for the future spatial
missions devoted to fundamental physics, the remaining fluctuations can 
be calculated up to the ultimate quantum level \cite{Courty00}. 

We now come back to the discussion of the Casimir force, {\it i.e.} the mechanical
force exerted by vacuum fluctuations on macroscopic mirrors \cite{Casimir48}.  
Casimir calculated this force in a geometrical configuration
where two plane mirrors are placed a distance $L$ apart from each other,
parallel to each other, the area $A$ of the mirrors being much larger than 
the squared distance. Casimir considered the ideal case 
of perfectly reflecting mirrors and obtained the following expressions for 
the force $F_{\rm Cas}$ and energy $E_{\rm Cas}$ 
\begin{equation}
F_{\rm Cas}  =  \frac{\hbar c \pi^2 A}{240 L^4}  \qquad \qquad
E_{\rm Cas}  =  \frac{\hbar c \pi^2 A}{720 L^3}  
\qquad \qquad  \left( A \gg L^2 \right)
\label{eqCasimir}
\end{equation}
This faint attractive force ($\sim 0.1 \mu$N for $A=1 {\rm cm}^2$ and 
$L=1\mu$m) has been observed in a number of experiments 
\cite{Deriagin57,Spaarnay58,Tabor68,Black68,Sabisky73,Blokland78}.
The accuracy in the first ``historical'' experiments was of the order of 
100 \%~: according to \cite{Spaarnay58} for example, the experimental results 
did ``not contradict Casimir's theoretical prediction''. 
It has been greatly improved in recent measurements 
\cite{Lamoreaux97,Mohideen98,Roy99,Harris00,Ederth00,Chan01} and this is 
important for at least two reasons.

First, the Casimir force is the most accessible experimental consequence of 
vacuum fluctuations in the macroscopic world. Due to the difficulties with 
vacuum energy evoked above, it is of course crucial to test with great care 
the predictions of Quantum Field Theory. Any pragmatic definition of 
vacuum necessarily involves a region of space limited by some enclosure and
the Casimir force is nothing but the physical manifestation of vacuum when 
it is enclosed in this cavity. Vacuum fluctuations are thus modified and 
vacuum energy depends on the distance $L$ so that mirrors are attracted 
to each other. In the ideal case of perfectly reflecting mirrors in vacuum,
the force only depends on the distance and on two fundamental constants, 
the speed of light $c$ and Planck constant $\hbar$. This is a remarkably 
universal feature in particular because the Casimir force is independent of the 
electronic charge in contrast to the Van der Waals forces. In other words,
the Casimir force corresponds to a saturated response of the mirrors which
reflect 100 \% of the incoming light in the ideal case, but cannot reflect 
more than 100 \%. However most experiments are performed at room temperature 
with mirrors which do not reflect perfectly all field frequencies and this
has to be taken into account in theoretical estimations
\cite{Lifshitz56,Mehra67,Brown69,Schwinger78}. 

Then, evaluating Casimir force is a key point in a lot of very accurate force 
measurements in the range between nanometer and millimeter. These experiments 
are motivated either by tests of Newtonian gravity at short distances 
\cite{Fischbach98} or by searches for new short range weak forces predicted 
in theoretical unification models \cite{Carugno97,Bordag99,Fischbach99,Long99}.
Basically, they aim at putting limits on deviations from present standard
theory through a comparison of experimental results with theoretical
expectations. Casimir force is the dominant force between two neutral 
non-magnetic objects in the range of interest and, since a high accuracy is needed,
it is therefore important to account for differences between the ideal case considered 
by Casimir and real situations studied in experiments. For theory-experiment 
comparisons of this kind in fact, the accuracy of theoretical calculations 
becomes as crucial as the precision of experiments \cite{Lambrecht00c}.

Let us first discuss the effect of imperfect reflection by introducing scattering 
amplitudes which depend on the frequency of the incoming field and obey general 
properties of unitarity, high-frequency transparency and causality. 
The Casimir force can thus be given a regular expression, free from the divergences 
usually associated with the infiniteness of vacuum energy \cite{Jaekel91}
\begin{equation}
F = \frac{A}{c^3} \sum_{\rm p}
\int \frac{{\rm d}\omega}{2\pi} \omega^2 \frac{{\rm d} \cos\theta}{2\pi} 
\cos^2\theta \ \hbar \omega  \left( 1 - g_{\rm p} \right) 
\qquad \qquad 
g_{\rm p}  = \frac{1- \left| r_{\rm p}\left(\omega , \kappa \right) \right|^2}
{\left| 1-r_{\rm p} \left(\omega , \kappa \right) e^{2i\kappa L} \right|^2}
\label{eqCasimirReal}
\end{equation}
The force is an integral over field modes of vacuum radiation pressure
effect with a sum over the two polarization states `p'~: $\hbar \omega$ 
represents the energy per mode, $\cos^2\theta $ an incidence factor with $\theta$ 
the incidence angle; the parenthesis shows that the Casimir force is the difference
between the radiation pressures on outer and inner sides of the cavity with the 
factor 1 representing outer side while the Airy function $g_{\rm p}$ describes the 
transformation of field energy from the outer to the inner side. The Airy function 
depends on the product $r_{\rm p} = r_{1,{\rm p}} r_{2,{\rm p}}$ of the reflection amplitudes 
associated with the two mirrors at a given frequency $\omega$ and longitudinal 
wavevector $\kappa = \frac \omega c \cos\theta$. The ideal Casimir result is
recovered at the limit where mirrors may be considered as perfect over the 
frequency range of interest, that is essentially over the first few resonance 
frequencies of the cavity. Since metals are perfect reflectors at frequencies 
lower than their plasma frequency, deviation from the ideal
result (\ref{eqCasimir}) becomes significant at distances $L$ shorter than a few
plasma wavelengthes, typically $L \leq \ \sim 0.3 \mu{\rm m}$ for gold or
copper \cite{Lambrecht00}.

A second important correction is due to the radiation pressure of thermal 
fluctuations which are superimposed to vacuum fluctuations as soon as the 
temperature differs from zero. The expression (\ref{eqCasimirReal}) of the force 
has then to be modified by replacing the vacuum energy $\hbar \omega$ per mode 
by the thermally modified energy 
$ \hbar \omega \times \coth \frac{\hbar \omega}{2 k_{\rm B} T}$.
The modification becomes large at low frequencies so that the thermal
deviation from the ideal formula is significant at large distances,
typically $L \geq \ \sim 3 \mu{\rm m}$ at room temperature \cite{Genet00}.

\begin{figure}[h]
\centerline{\psfig{figure=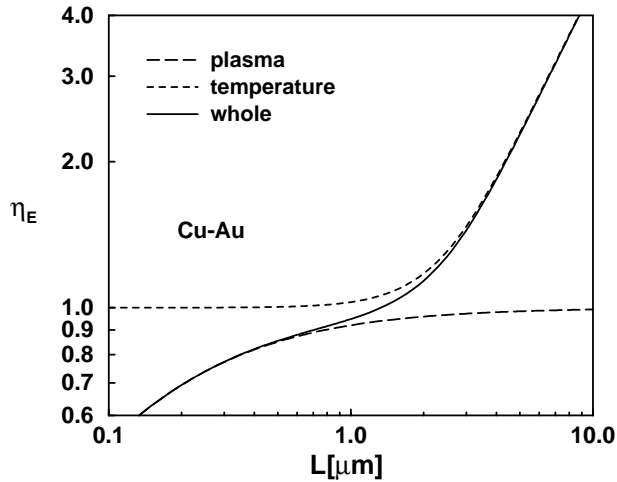,height=7cm}}
\caption{Variation of the factor $\eta_{\rm E} = \frac E E_{\rm Cas} $ as a function
of distance $L$; the long-dashed and short-dashed curves respectively represent the 
corrections associated with imperfect reflection and non zero temperature; the solid 
curve is the whole correction; these curves have been evaluated for a plasma model 
with the plasma wavelength corresponding to copper or gold.}
\label{figCorr}
\end{figure}

The corrections to the ideal Casimir energy (\ref{eqCasimir}) are shown 
on Figure \ref{figCorr} where we have plotted the factors 
$\eta_{\rm E} = \frac E E_{\rm Cas} $
corresponding respectively to imperfect reflection and to temperature correction
as well as the whole correction when both effects are taken into account. 
Since the two corrections are significant in non overlapping distance ranges, 
the whole correction is approximately the product of the 
two effects evaluated separately \cite{Lambrecht01}. 

This kind of theoretical evaluations should make possible accurate comparison, 
say at the 1\% level, between experiment and theory, provided other discrepancies 
between ideal and real situations are also mastered. Recent experiments have not 
been performed in the configuration of two plane plates but of a sphere and a plane.
The Casimir force in this geometry is usually estimated from the proximity theorem 
\cite{Deriagin68}. Basically this amounts to obtaining the force as the integral 
of contributions of the various inter-plate distances as if they were independent. 
In plane-sphere geometry, the force evaluated in this manner is just given by the 
Casimir energy obtained in plane-plane configuration for the distance $L$ of closest 
approach in plane-sphere geometry. Then, the factor $\eta_{\rm E}$ drawn on Figure 
\ref{figCorr} is used to infer the factor for the force measurement in plane-sphere 
geometry. The accuracy of this approximation however remains to be mastered in a more 
reliable manner. This is also the case for surface roughness corrections which play 
an important role for short distance measurements \cite{Bezerra97}. 

Clearly, these points have to be inspected with great care before any claim for an 
accurate agreement between theory and experiments. It is however clear that the 
existence of the Casimir force, its sign and magnitude have now been experimentally 
demonstrated, to within a few \% for the magnitude. Hence, it is certainly no longer 
possible to dismiss the mechanical effects of vacuum. Then the question arises of the 
consistency of these effects with the principles of relativity theory.
As argued above, it is possible to formulate well-defined questions and to answer at 
least some of them. Let us discuss here one such question which is directly connected 
to the question of the validity of Einstein equivalence principle.

As already discussed, even if the mean value of vacuum energy does not contribute to 
gravity, energy differences have to contribute. This is in particular the case for 
the Casimir energy, a variation of the vacuum energy with the length of the cavity. 
If Einstein equivalence principle is obeyed for such energy contributions, Casimir 
energy should also contribute to the inertia of the Fabry-Perot cavity. Clearly, 
this effect has a small influence on the motion of a macroscopic cavity but is of 
fundamental importance, since it is a quantum version, at the level of vacuum 
fluctuations, of Einstein argument for the inertia of a box containing a photon 
bouncing back and forth \cite{Einstein05,Einstein06}.
Here, Einstein law of inertia of energy has to be applied to
the case of a stressed body, where it is read \cite{Einstein07} 
\begin{equation}
F_{\rm mot} = - \mu a \qquad \qquad \mu = \frac {E_{\rm Cas} - F_{\rm Cas} L} {c^2}
\end{equation}
A detailed investigation of this question requires an evaluation of Casimir forces
 when the mirrors of the cavity are allowed to move \cite{Jaekel92JP}. When specifying 
a global motion of the cavity with uniform acceleration, the obtained force perfectly 
fits the contribution of Casimir energy to inertia \cite{Jaekel93JPb}.
This confirms that variations of vacuum energy effectively contribute to gravitation 
and inertia as expected from the general principles of relativity.

Going further along the same lines, fluctuations of vacuum energy should also contribute 
to gravitation. This implies that quantum fluctuations of stress tensors and of spacetime 
curvatures are necessarily coupled to each other 
\cite{deWitt62,Feynman63,Weinberg65,Zeldovich86}. This problem can be treated by linear
response techniques \cite{Jaekel95AP}, in analogy with the study of fluctuations of 
moving objects coupled to the quantum fluctuations of forces acting upon them
\cite{Jaekel97}. The latter study has important consequences for the question
of relativity of motion \cite{Jaekel98}. 

Even a single mirror at rest in vacuum is submitted to the effect of vacuum radiation 
pressure \cite{Jaekel92QO,Barton94}. The resulting force has a null mean value due to 
the balance between the contributions of opposite sides but still has fluctuations 
since the vacuum fluctuations impinging on both sides are statistically uncorrelated. 
When the mirror is moving, the balance is also broken for the mean force which
entails that vacuum exerts a radiation reaction against motion. The dissipative force 
is described by a susceptibility allowing one to express the force $F_{\rm mot}$ 
as a function of motion $q$, in a linear approximation and in the Fourier domain, 
\begin{equation}
F_{\rm mot}[\Omega] = \chi [\Omega]   q[\Omega]
\end{equation}
The motional susceptibility $\chi [\Omega]$ is directly related to the force fluctuations 
evaluated for a mirror at rest through quantum fluctuation-dissipation relations. 
The damping of mechanical energy is associated with an emission of radiation and the 
radiation reaction force is just the consequence of momentum exchange. 

Let us first consider the simple case of a mirror moving with a uniform velocity. 
When this motion takes place in vacuum, the radiation reaction force vanishes, so that 
the reaction of vacuum cannot distinguish between inertial motion and rest, 
in full consistency with the principle of relativity of motion. 
But a friction force arises when a mirror moves in a thermal field, in analogy with 
the damping of motion by air molecules \cite{Einstein05B,Einstein17}. 
With the notations of the present paper, this force corresponds for a perfectly 
reflecting mirror in the limits of large plane area and large temperature to 
\begin{equation}
\chi [\Omega] \simeq \frac {i \hbar A}{240 \pi^2 c^4} \theta^4 \Omega    
\qquad \qquad
F_{\rm mot} \simeq  \frac {\hbar A}{240 \pi^2 c^4} \theta^4 q^{\prime}(t)
\qquad \qquad
\left( A \gg \frac{c^2}{\Omega^2} \ ,\ \theta \gg \Omega \right)
\label{eqMotionalHighT}
\end{equation}
This expression vanishes when $\theta$ is set to zero, but this does not lead to 
the absence of any dissipative effect of vacuum on a moving scatterer. For
a perfect mirror with an arbitrary motion in electromagnetic quantum vacuum, 
we obtain a susceptibility proportional to the fifth power of frequency or, 
equivalently, a force proportional to the fifth order time derivative of
the position
\begin{equation}
\chi [\Omega] \simeq \frac {i \hbar A}{60 \pi^2 c^4} \Omega^5  
\qquad \qquad F_{\rm mot} \simeq - \frac {\hbar A}{60 \pi^2 c^4}   
q^{\prime \prime \prime \prime \prime}(t)
\qquad \qquad
\left( A \gg \frac{c^2}{\Omega^2} \ ,\ \theta = 0 \right)
\label{eqMotionalZeroT}
\end{equation}
This result can be guessed from the previous one through a dimensional 
analysis~: the factor $\theta^4$ appearing in (\ref{eqMotionalHighT}) 
has been replaced by $\Omega^4$ in (\ref{eqMotionalZeroT}). This 
argument played a similar role in the discussion of the expression
(\ref{eqVacuumEnergyDensity}) of vacuum energy density with however
an important difference~: the vacuum term is now perfectly regular 
and it corresponds to a well-defined physical effect. 

This effect was first analyzed in 1976 by Fulling and Davies for the simpler 
case of a perfectly reflecting mirror moving in vacuum of a scalar field theory 
in two-dimensional spacetime \cite{Fulling76} (see also \cite{deWitt75,Ford82,Birrell82}).
In this case, the dissipative force is proportional to the third order time 
derivative of the position since $\frac {A}{c^2} \Omega^5$ is replaced by $\Omega^3$ 
in the preceding dimensional argument. The vacuum reaction force has thus the same 
form as for an electron in electromagnetic vacuum and it raises the same causality 
and stability problems \cite{Rohrlich65}. This difficulty is however solved by taking 
into account the fact that any real mirror is certainly transparent at high frequencies. 
This leads to a completely satisfactory treatment of motion of a real mirror 
in vacuum \cite{Jaekel92PL}. Furthermore, the stable equilibrium reached by
the mirror's motion coupled to vacuum radiation pressure fluctuations
contains a consistent description of the quantum fluctuations of the 
mirror, generalizing the standard Schr\"{o}dinger equation \cite{Jaekel93JPa}.

We may emphasize that the motional force does not raise any 
problem to the principle of special relativity. As a matter of fact,
the reaction of vacuum (\ref{eqMotionalZeroT}) vanishes in the particular
case of uniform velocity. The quantum formalism gives an interesting 
interpretation of this property~: vacuum fluctuations appear exactly 
the same to an inertial observer and to an observer at rest. 
Hence the invariance of vacuum under Lorentz transformations is an essential 
condition for the principle of relativity of motion to be valid and it 
establishes a precise relation between this principle and the symmetries of vacuum.
More generally, vacuum does not oppose to uniformly accelerated motions and this 
property corresponds to conformal symmetry of quantum vacuum \cite{Jaekel95QSO}. 
In this sense, vacuum fluctuations set a class of privileged reference frames 
for the definition of mechanical motions.

At the same time, the existence of dissipative effects associated with 
motion in vacuum challenges the principle of relativity of motion in its more
general acceptance. Arbitrary motion produces observable effects, namely the 
resistance of vacuum against motion and the emission of radiation by the 
moving mirror, although there is no further reference for this motion
than vacuum fluctuations themselves. This means that quantum theory has built up 
a theoretical framework where the questions raised in the introduction may
find consistent answers. The space in which motion takes place can no longer 
be considered as empty since vacuum fluctuations are always present. 
These fluctuations give rise to real dissipative effects for an arbitrary motion. 
However these effects vanish for a large category of specific motions,
including the cases of uniform velocity and uniform acceleration.

Clearly, it would be extremely interesting to obtain experimental evidence for 
the dissipative effects associated with motion in vacuum \cite{Davies96}. 
These effects are exceedingly small for any motion which could be achieved 
in practice for a single mirror, but an experimental observation is conceivable
with a cavity oscillating in vacuum. In this case, the emission of motional 
radiation is resonantly enhanced \cite{Lambrecht96}, specific signatures are
available for distinguishing the motional radiation from spurious effects
\cite{Lambrecht98} so that an experimental demonstration appears to be 
achievable with very high finesse cavities \cite{Jaekel01}.

\noindent {\bf Acknowledgements}\\
We thank E.~Fischbach, J.~Long and R.~Onofrio for stimulating discussions.

\end{document}